\documentclass[twocolumn]{aastex63}

\usepackage{wrapfig}
\usepackage{tcolorbox}
\usepackage{stix}
\usepackage{appendix}
\usepackage[utf8]{inputenc}
\usepackage{floatflt}
\usepackage{lineno}
\graphicspath{{./}{figures/}}

\received{2023 November 7}
\revised{2023 December 23}
\accepted{2024 April 13}
\submitjournal{PSJ}
\begin{document}

\title{Laser Irradiation of Carbonaceous Chondrite Simulants: Space Weathering Implications for C-complex Asteroids}

\correspondingauthor{Andy J. L\'{o}pez-Oquendo}
\email{al2987@nau.edu}

\author[0000-0002-2601-6954]{Andy J. L\'{o}pez-Oquendo}
\affiliation{Department of Astronomy and Planetary Science, Northern Arizona University, Flagstaff, AZ 86011, USA}

\author[0000-0002-1551-3197]{Mark J. Loeffler}
\affiliation{Department of Astronomy and Planetary Science, Northern Arizona University, Flagstaff, AZ 86011, USA}
\affiliation{Center for Material Interfaces in Research and Applications, Northern Arizona University, Flagstaff, AZ 86011, USA}

\author[0000-0003-4580-3790]{David E. Trilling}
\affiliation{Department of Astronomy and Planetary Science, Northern Arizona University, Flagstaff, AZ 86011, USA}


\begin{abstract}

Surfaces of carbonaceous asteroids (C-complex) have shown diverse contrasting spectral variations, which may be related to space weathering. We performed laser irradiation experiments on CI and CM simulant material under vacuum to mimic the spectral alteration induced by micrometeorite impacts. We used in situ ultraviolet-visible and near-infrared reflectance spectroscopy to analyze spectral alterations in response to pulsed laser irradiation, as well as scanning electron microscopy and x-ray photoelectron spectroscopy to search for microstructural and compositional changes. Laser irradiation causes an increase in spectral slope (reddening) and a decrease in the albedo (darkening), and these changes are stronger in the ultraviolet-visible region. These spectral changes are likely driven by the excess iron found in the altered surface region, although other factors, such as the observed structural changes, may also contribute. Additionally, while the 0.27~${\mu}m$ band appears relatively stable under laser irradiation, a broad feature at 0.6~${\mu}m$ rapidly disappears with laser irradiation, suggesting that space weathering may inhibit the detection of any feature in this spectral region, including the 0.7~${\mu}m$ band, which has typically been used an indicator of hydration. Comparing our laboratory results with optical spectrophotometry observations of C-complex asteroids, we find that the majority of objects are spectrally red and possess colors that are similar to our irradiated material rather than our fresh samples.  Furthermore, we also find that ``younger'' and ``older'' C-complex families have similar colors, suggesting that the space weathering process is near-equal or faster than the time it takes to refresh the surfaces of these airless bodies.  

\end{abstract}


\section{Introduction \label{sec:intro}}

The combination of exogenic phenomena such as micrometeorite impacts (interplanetary dust) and cosmic and solar wind radiation are known to successfully modify the surface of airless planetary bodies. These chemical and physical changes occurring within individual or conglomerates of particles are commonly known as space weathering \citep{Hapke2001, Chapman2004}. The compositional and structural alteration induced by space weathering (SW) affects our interpretation of remote sensing characterization and, by extension, the efforts to decode early solar system conditions from meteorites and returned asteroid samples \citep{Yada2022}. Thus, it is critical to understand how these processes modify the surface of any airless body we hope to fully characterize.

The causes (formation of chemically reduced npFe/Fe$^{0}$ particles) and consequences (spectral reddening and darkening) of space weathering on silicate-rich minerals (i.e., ordinary chondrite meteorites, S-type asteroids, and the Moon) in the visible to near-infrared region are well-understood at both optical, spectral, and chemical levels \citep{Yamada1999, Sasaki2001, Clark2002, SASAKI20032537, Chapman2004, Hapke2001, BRUNETTO2006546, Loeffler2009, Thompson2016}. In contrast to the silicate-rich bodies, the effects of SW on carbonaceous-containing bodies (i.e., C-type asteroids and carbonaceous chondrites meteorites) is still under debate. 
 
Many pulsed-laser irradiation experiments have been conducted to simulate space weathering induced by micrometeorite impacts in several carbonaceous chondrite meteorite samples. The number of studies centered on understanding the impact-induced space weathering on carbonaceous meteorites (i.e., Murchison (CM2), Allende (CV3), Orgueil-Ivuna (CI1), Tagish Lake (C2), Yamato (CM2), ElQuss Abu Said (CM), and others), has resulted in a wide range and, in some cases, possibly conflicting results. For instance, these analyses have shown: 1) initial spectral bluing (decrease in spectral slope) \citep{Hiroi2004, Hiroi2013, Gillis-Davis2015, Matsuoka2015,Zhang2022}, 2) spectral reddening (increase in spectral slope) \citep{Thompson2019, GILLISDAVIS20171, Moroz2004a, MOROZ1996366, Brunetto2014}, 3) competition between bluing and reddening with progressive laser irradiation \citep{GILLISDAVIS20171, Kaluna2017}, 4) depletion of hydrated spectral features in the visible range \citep{Kaluna2017, Matsuoka2015}, and$/$or 5) no evident spectral slope alteration \citep{Vernazza2013, gillis2018laser, Kaluna2017} but absorption band suppression observed in one case \citep{Kaluna2017}. In addition, the optical brightness of irradiated meteorite samples has also shown contrasting albedo trends within the same meteorite (i.e., Tagish Lake \citep{Hiroi2004, Hiroi2013} and Murchison \citep{Matsuoka2015, Matsuoka_2020}). It is possible that these spectral mismatches arise from mineralogical variabilities within the sample or even possibly the degree to which the sample has been weathered \citep{GILLISDAVIS20171, Kaluna2017, Lantz2017}, suggesting that SW of carbonaceous asteroids is a fairly complex process.

From an observational perspective, a few works have studied the spectral trends among carbonaceous asteroids to understand space weathering consequences on their surfaces. \citet{Nesvorny2005} investigated space weathering in the main belt using the Sloan Digital Sky Survey (SDSS). They found a possible correlation between principal component analysis of reflectance spectra and families' ages, leading to the conclusion that C-complex asteroids experience a decrease in spectral slope over time (spectral bluing). Conversely, \cite{Lazzarin2006} used the Small Main-Belt Asteroid Spectroscopic Survey (SMASS) and found that the surfaces of dark asteroids (i.e., carbon-containing or C-types) became redder under the exposure of space weathering. \citet{Lantz2013} used the SMASS catalog and reflectance spectra of CM meteorites \citep{CLARK2011462} from the Brown University Reflectance Laboratory (RELAB) and found a similar trend as \citet{Nesvorny2005}. In recent work with the SDSS catalog, \citet{Thomas2021} studied how the spectral slope changed as a function of object size across different carbon-containing families. While these authors found possible trends for some families (i.e., reddening on the Themis family, consistent with \citet{Kaluna2016}, and blueing on the Hygiea family), a general trend could not be established. As stated by \citet{Lantz2017}, \citet{Thomas2021} speculated that the different spectral trends could be caused by variations in initial composition on individual carbonaceous asteroid surfaces (i.e., aqueous alteration or abundance of organic materials). 

\citet{LANTZ201810} used spectra of 34 primitive Near-Earth Asteroids (NEAs) from various spectral catalogs such as the MIT Hawaii Near-Earth Objects Spectroscopic Survey (MITHNEOS), the SDSS, and the SMASS. They compared the asteroid's spectra with the ion-irradiated meteorites presented in \citet{Lantz2017} to understand their taxonomical evolution by space weathering and found that numerous carbonaceous chondrites meteorites show small or no spectral change with space weathering. As summarized in their table 3, asteroids spectrally correlated with CO/CV meteorites are likely to experience reddening and a concave-like spectral curvature between the visible to near-infrared wavelengths, while those sharing similarities with the CI/CM/Tagish Lake meteorites are more susceptible to undergo spectral bluing and a convex spectral curvature in their spectra.  This is reasonably consistent with other ion irradiation studies of meteorites, which have shown Allende and FM 95002 redden and darken \citep{Lazzarin2006, Brunetto2014}, while Tagish Lake meteorite showed a decrease in spectral slope with Ar$^{+}$ ion irradiation \citep{Vernazza2013}, although no changes were observed with less damaging He$^{+}$ ions.

Piecing together the laboratory experiments and observational evidence of space weathering on carbonaceous surfaces shows that the space weathering behavior on C-complex asteroids is complicated. However, it is possible that spectral changes induced by simulated micrometeorite impacts may follow the compositional dependence proposed for ions in \citet{Lantz2017} and \citet{LANTZ201810}. Thus, in an attempt to test this possibility and to better understand space weathering effects induced by micrometeorite impacts into carbon-rich regoliths, we have performed pulsed laser irradiation experiments on CI and CM carbonaceous chondrite simulant material \citep{Britt2019}. We selected simulant powder instead of a real meteorite sample as simulant materials are available in large quantities and allow us to check the reproducibility of our results. These simulants are also spectrally similar to aqueously altered carbonaceous chondrite meteorite \citep{CLOUTIS2011309, Britt2019} and possibly asteroid Bennu \citep{CLARK2011462, Hamilton2019}. During irradiation, we monitored the spectral reflectance of our samples in the ultraviolet-visible to near-infrared region, and after irradiation, we investigated changes in surface morphology with electron microscopy and surface composition with X-ray photoelectron spectroscopy. We quantified changes in the spectral slope and albedo in these spectral regions so that we could make a direct comparison with visible spectrophotometry observations found in the SDSS catalog.  This novel comparison approach allows us to explore to what degree the observed laboratory spectral trends may be extended to explain space weathering effects on C-complex asteroids.

\section{Materials and Methods \label{sec:method}}

\subsection{Sample preparation, Laser irradiation, and In Situ Analysis}

In all experiments, we used CI and CM simulant powder from the Center of Lunar and Asteroid Surface Science at the University of Central Florida \citep{Britt2019}. To prepare a sample, we first dry-sieved the powder to a grain size of $<$45~${\mu}m$. As in \citet{Prince2022}, we opted to use the smaller grain size, compared to the 45–125~${\mu}m$ size fraction we have used previously for laser irradiation \citep{Loeffler2016, Prince2020}, because prior characterization of these simulant grains shows that material larger than 50~${\mu}m$ are typically agglomerates of smaller grains \citep{Britt2019}. Next, we placed about 85~mg of powder into a 12~mm aluminum ring and pressed it into a pellet by compressing it between two stainless steel disks with a 10~ton load for $\sim$5~min. The resulting pellet was $\sim$10~mm in diameter. Finally, we mounted the pellet onto an aluminum sample holder and inserted it into an ion-pumped ultrahigh vacuum chamber with a pressure of $<$5$\times10^{-8}$~Torr \citep{Prince2020}. We performed in-situ reflectance analysis of our sample before, during, and after laser irradiation using a combination of ultraviolet-visible (UV-VIS) and near-infrared (NIR) spectroscopy between 0.24 and 2.5~${\mu}m$. For all measurements, we aimed both light sources directly at the sample’s surface (0$^{\circ}$ incidence) and collected the reflected light at an angle of 30$^{\circ}$. We also never changed the orientation of our samples within the sample holder, as we had done for a few select samples in previous work \citep{Prince2020}. Unless noted otherwise, for laser irradiation and reflectance acquisition, we used the procedure and parameters described in our recent work \citep{Prince2020}.

\begin{figure}[ht!]
    \centering
    \includegraphics[width=0.48\textwidth]{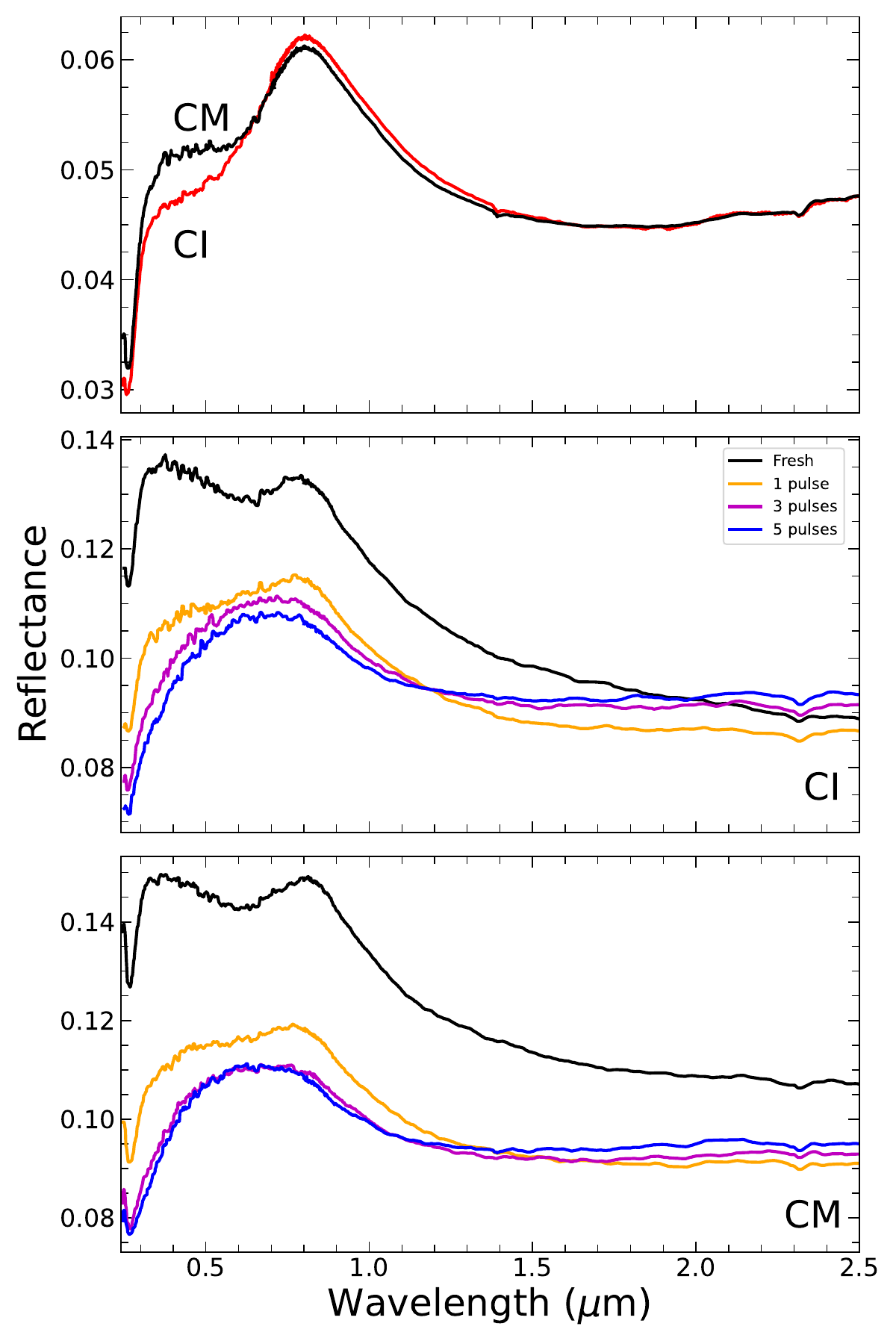}
    \caption{Absolute reflectance of CI and CM simulant material before and during laser irradiation between 0.24 and 2.5~${\mu}m$. Top: spectra of loose CI (red line) and CM (black line) simulant powder obtained under ambient conditions. Middle: CI spectra (from top to bottom at 0.3~${\mu}m$) correspond to the spectrum after 0 (black), 1 (orange), 3 (magenta), and 5 (blue) pulses. Bottom: CM spectra (from top to bottom at 0.3~${\mu}m$) correspond to the spectrum after 0 (black), 1 (orange), 3 (magenta), and 5 (blue) pulses. One laser pulse corresponds to one raster across the sample.}
    \label{fig:spectra}
\end{figure}

We developed a Python code to perform quantitative analysis of the spectral alteration induced by laser irradiation. We made use of the \textit{Python} package \textit{SciPy} \citep{scipy2020} to find the average spectral slope over the wavelength range of interest within 95\% confident intervals. For the ultraviolet-visible (UV-VIS) wavelength region, we measured the average spectral slope between 0.3 and 0.7~${\mu}m$. For the near-infrared (NIR), we measured the average spectral slope between 0.8 to 2.4~${\mu}m$. The average spectral slope is a convenient way to express the derivative of the reflectance when there is a non-linear behavior of the spectrum within the wavelength region (e.g., absorption features) and has often been used in previous works \citep{LOEFFLER2022114881}. We also quantify changes in the albedo of the sample during irradiation by reporting the reflectance at 0.55, 1, and 2~${\mu}m$, as well as the relative change of the normalized reflected intensity at these wavelengths ($R(\lambda)$/$R(\lambda)_{0}-1$), where ($R$) is the reflectance of the weathered sample and ($R_{0}$) is the reflectance of the fresh sample.

\subsection{Ex-situ analysis}

\subsubsection{Scanning Electron Microscope}
We performed scanning electron microscope (SEM) imaging on selected CI and CM samples using a Zeiss SUPRA 40VP field emission electron microscope located in the Imaging and Histology Core Facility at Northern Arizona University. We acquired secondary electron (SE) images of the samples in high vacuum mode using an accelerating voltage of 5~kV and line average noise reduction (N=3). These parameters allowed us to minimize surface charging of our sample without having to coat the surface with another material (e.g., carbon). We imaged both unirradiated and irradiated areas at similar magnifications to directly compare microstructural alterations induced by laser irradiation.

\subsubsection{X-Ray Photoelectron Spectroscopy}

We also performed ex-situ X-ray Photoelectron spectroscopy (XPS) using a PHI 5600 surface analysis system equipped with a Mg K$\alpha$ X-ray source (1253.6 eV). While surface reactions can occur during atmospheric exposure, masking changes in chemical composition \citep{Loeffler2009, Loeffler2016}, ex-situ analysis can still be useful to determine if the elemental abundances in the deposit are significantly different from the starting material. Thus, we acquired survey and high-resolution spectra for selected fresh and laser-irradiated CM samples. We took survey spectra using 1~eV steps with a dwell time of 50~ms per step, a pass energy of 117.4~eV ($\sim$1.8~eV energy resolution), and then averaged over 10 scans. We acquired high-resolution data in 0.2~eV steps with a dwell time of 50~ms, a pass energy of 46.95~eV ($\sim$0.7~eV energy resolution) and then averaged over 21 scans for C-1s, Mg-2p, Na-1s, Si-2p, 42 scans for Fe-2p, and 7 scans for O-1s features. We calibrated the binding energy scale to adventitious carbon (C-1s) at 284.8~eV \citep{POWELL1979361}. We performed spectral fitting by utilizing CASA XPS software \citep{FAIRLEY2021100112} and used a Shirley model to remove the inelastic background from the photoelectron spectra \citep{Shirley1972}.

\begin{figure}[ht!]
    \centering
    \includegraphics[width=0.47\textwidth]{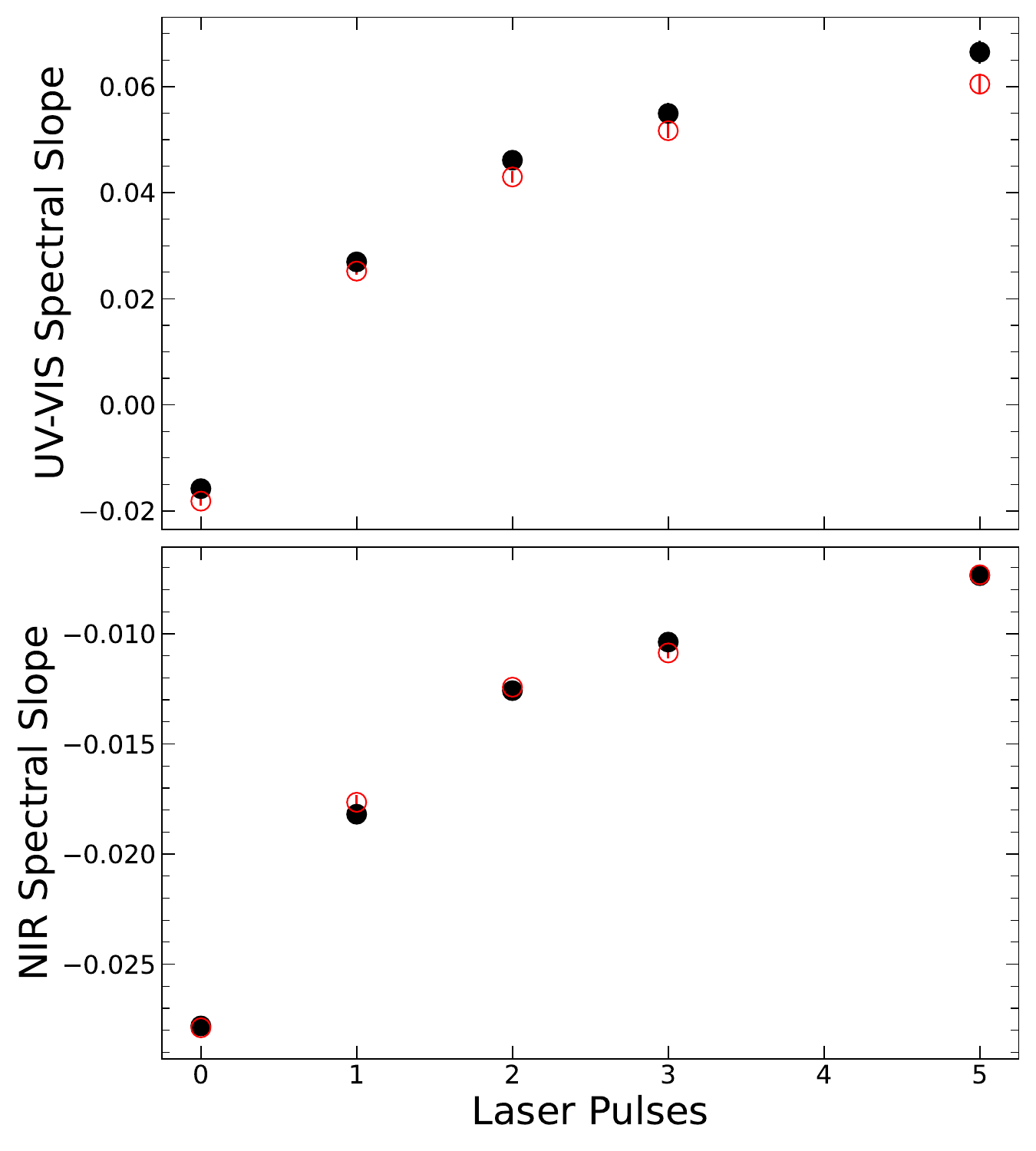}
    \caption{Changes in the average spectral slope for the CI (empty red symbols) and CM (filled black symbols) sample during laser irradiation. Top: average spectral slope between 0.3 and 0.7~${\mu}m$. Bottom: average spectral slope between 0.8 and 2.4~${\mu}m$.}
    \label{fig:Vis_slope}
\end{figure}

\begin{figure}[ht!]
    \centering
    \includegraphics[width=0.47\textwidth]{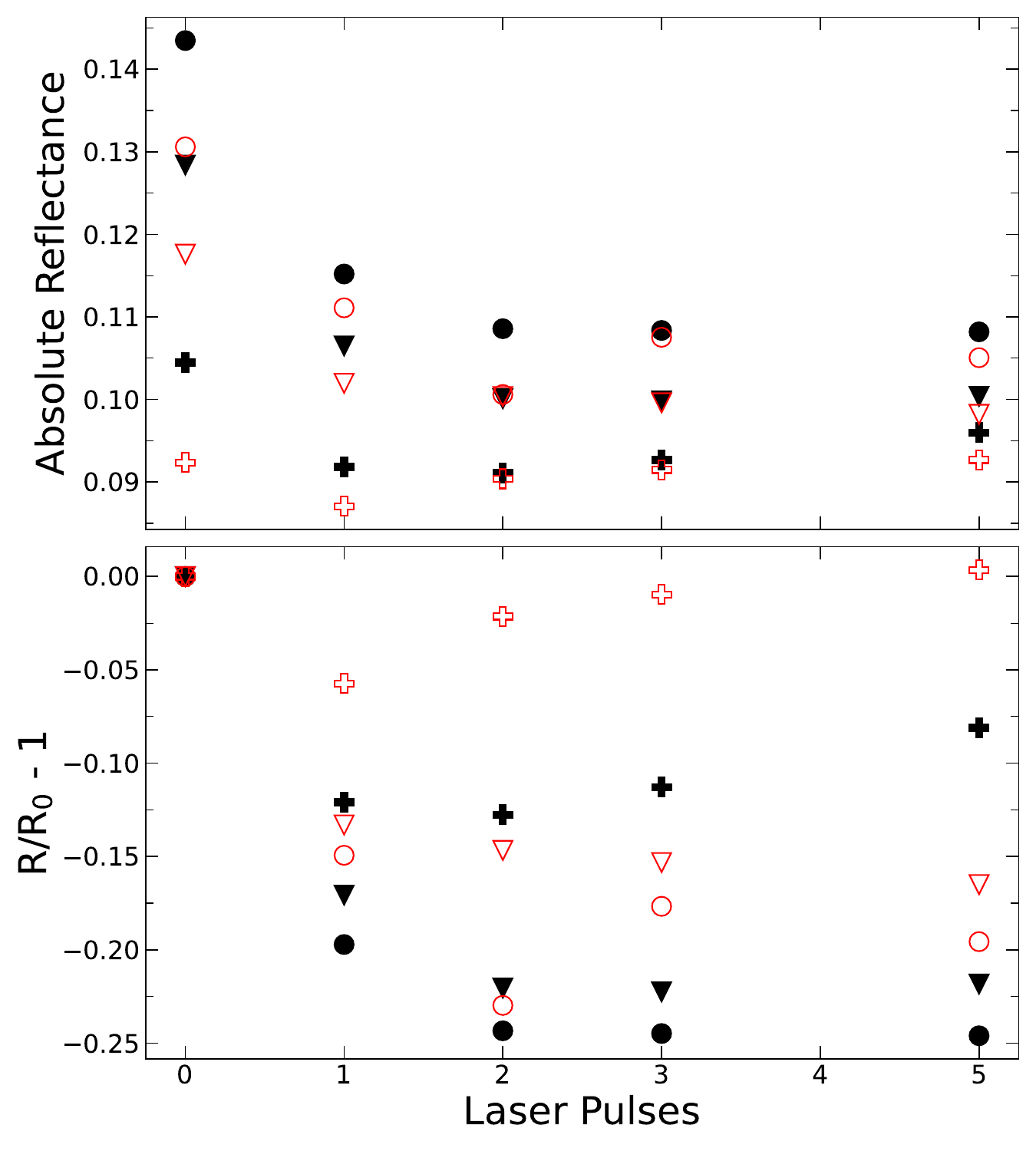}
    \caption{Absolute and relative change of the} normalized reflected intensity for the CI (empty red symbols) and CM (filled black symbols) sample during laser irradiation. Top: absolute reflectance at 0.55~${\mu}m$ (circles), 1.0~${\mu}m$ (inverted triangles), and 2.0~${\mu}m$ (crosses). Bottom: normalized reflected intensity at 0.55~${\mu}m$ (circles), 1.0~${\mu}m$ (inverted triangles), and 2.0~${\mu}m$ (crosses).
    \label{fig:brightness}
\end{figure}

\subsection{Comparing laboratory spectroscopy with spectrophotometric surveys}

\subsubsection{Laboratory spectra optical colors}
The large spectrophotometric surveys, which contain measured colors of thousands of asteroids, are a powerful tool to investigate the surface properties of rocky airless bodies in our Solar System. In order to compare our laboratory results to these surveys, we converted the spectral reflectance to magnitude (M$_{f}$) at a given bandpass filter ($f$) using a similar spectral convolution approach as in \citet{Mommert2016} and \citet{Lopez-Oquendo_2022}:
\begin{equation}
 M_{f} = -2.5 \log_{10}{\left( \frac{\int R(\lambda)A(\lambda)\lambda d\lambda}{\int R(\lambda)\lambda d\lambda} \right)} 
 \label{eq:convolution}
\end{equation}
where A($\lambda$) is the laboratory spectrum, R($\lambda$) is the filter response function for the Sloan Digital Sky Survey (SDSS; \citet{Gunn1998}), and $\lambda$ is the wavelength region evaluated in the convolution. After converting the reflectance spectra in Figure \ref{fig:spectra} to relative magnitude in each $g$ and $r$ bandpasses, we calculated the $g-r$ color. To properly correlate the color of laboratory spectra to the measured color of asteroids, one must simulate the identical conditions as if our laboratory pellet were in space reflecting sunlight. To address this, we took the Sun magnitudes at the SDSS $g$ and $r$ filters \citep{Willmer2018} and computed the Sun $g-r$ color. Finally, as asteroid magnitudes in the SDSS catalog are calibrated to the AB photometric system, we corrected the laboratory $g-r$ color by adding the $g-r$ color of the Sun in the AB system to produce the apparent color that our laboratory spectrum would have if observed in space.

\subsubsection{C-complex families optical colors}

To directly compare the irradiated sample colors with observed asteroid colors, we used the fourth release of the SDSS Moving Object Catalog (MOC, \citet{Ivezic2004}), which contains 471,569 moving objects or about 13 times more than was available in previous work \citep{Nesvorny2005}. From this SDSS data, we selected objects in C-complex asteroid families using the Nesvorny dynamical family catalog, version~3.0 \citep{Nesvorny2015}. For this analysis, we established several criteria to remove low-quality SDSS MOC data. We used magnitude cutoffs of 22.2, 22.2, 21.3, and 20.5~mag for the $g$-, $r$-, $i$-, and $z$-band in order to assure a 95\% photometric accuracy on each of these bands \citep{Ivezic, Demeo2013}. We excluded the u-band because of the significantly higher errors in the data (see Figure 2 in \citet{Demeo2013}), mostly attributed to the lower quantum efficiency of this filter compared to other filters coupled with the high atmospheric extinction at UV wavelengths \citep{Fukugita1996}. Furthermore, we removed poor photometric data (both entire nights and individual measurements) following \citet{Thomas2021} and \citet{Demeo2013}, using these data quality flags: \textit{badsky}, \textit{binned4}, \textit{deblend\_degenerate}, \textit{notchecked}, \textit{nodeblend}, \textit{stationary}, \textit{peaks\_too\_close}, \textit{edge}, \textit{bad\_moving\_fit}, and \textit{too\_few\_good\_detections}. These flags eliminate possible false associations with moving objects, sources that cannot be deblended, objects detected only in the 4$\times$ binned frame, measurements with the poorly determined local sky, and data taken too close to the edge of the frame. A detailed description of these flags can be found on the SDSS data release webpage\footnote{\url{https://classic.sdss.org/dr4/products/catalogs/flags.html}} \citep{Ivezic}. In addition, we restricted our analysis to seven carbonaceous families, which contained photometry data for more than 50 unique objects: Hygiea (289), Themis (383), Ursula (123), Veritas (88), Adeona (185), Dora (104), and Erigone (93). Finally, we computed the $g-r$, $r-i$, and $i-z$ colors for each member in the families, though here we only discuss the $g-r$ color as these filters cover the most relevant visible wavelengths (i.e., from 370~nm to 720~nm). We also obtained the family dynamical ages of each family from \citet{MARZARI1995168}, \citet{Nesvorny2003}, \citet{Nesvorny2005}, \citet{Broz2013}, and \citet{Nesvorny2015a} to find a possible correlation with space weathering and timescales.

\section{Results \label{sec:results}}

\subsection{Spectral Changes Induced  by Laser Irradiation \label{sec:spectra_results}}

In Figure \ref{fig:spectra}, we show the reflectance spectra of loose powders of fresh CI and CM simulant material, as well as their pellets before and after a pre-specified number of laser pulses. The spectra of the simulant powders (top panel of Figure \ref{fig:spectra}) are of low albedo, consistent with a characteristic property of dark carbonaceous asteroids \citep{Masiero2014}. The reflectance spectra of the CI and CM pellets are overall brighter than the loose powder, which is a consequence of the additional specular properties of the surface after pressing the pellet. Another effect of pressing the material is that it preferentially increases the reflectance at low wavelengths, causing the initial slope to appear blue in the ultraviolet-visible and near-infrared regions, and enhances a broad and weak feature centered at $\sim$0.6~${\mu}m$. 

The spectra of the fresh samples are consistent with those previously reported for loose powder samples \citet{Britt2019}. Both CI and CM spectra are also similar to those reported for the CM2 Murchison \citep{BECK2018124, Matsuoka_2020, Thompson2019} and CI1 Orgueil meteorites \citep{Bland2004}. As illustrated in Figure \ref{fig:spectra}, an absorption feature near 0.27~${\mu}m$ is present in the spectrum of the CI and CM simulant. It is likely that the UV band is due to the metal-O charge transfer from the octahedral Fe$^{2+}$ charge transfer, which is known to form an absorption band centered at 267~nm \citep{Tossell1974, Loeffler1974}. Interestingly, this feature was not identified in the initial characterization of these simulant materials \citep{Britt2019}, as their spectra did not extend below 0.4~${\mu}m$. The weak 2.3~${\mu}m$ absorption feature has been observed in other CM meteorites and attributed to the Mg-serpentines \citep{CLOUTIS2011309}. The peak at 0.8~${\mu}m$ has been attributed to maghemite contamination in the magnetite in the sample \citet{Britt2019}. 

Figure \ref{fig:spectra} also shows the response of our samples to laser irradiation. After a single laser pulse, the reflectance decreases, and the spectral slope increases (reddens) in both samples. In addition, the weak feature near 2.3~${\mu}m$ appears relatively unaffected by laser irradiation, while the 0.27~${\mu}m$ feature is harder to assess, as only being able to observe a portion of the band makes a robust quantitative analysis of this feature problematic. Interestingly, the broad feature at $\sim$0.6~${\mu}m$ is essentially removed after a single laser pulse. 

We quantify the trends in spectral slope and albedo for both the UV-VIS and NIR regions in Figure \ref{fig:Vis_slope}. The average spectral slope in the UV-VIS region increases by about a factor of four after laser irradiation, while the average spectral slope in the NIR goes from being spectrally blue to nearly flat, suggesting a reddening spectral effect. In addition, we find that the irradiated sample darkens by 25\% at 0.55~${\mu}m$,  $\sim$20\% at 1.0~${\mu}m$ and $\sim$10\% at 2~${\mu}m$ (Figure \ref{fig:brightness}).

\begin{figure*}[ht!]
    \centering
    \includegraphics[width=0.9\textwidth]{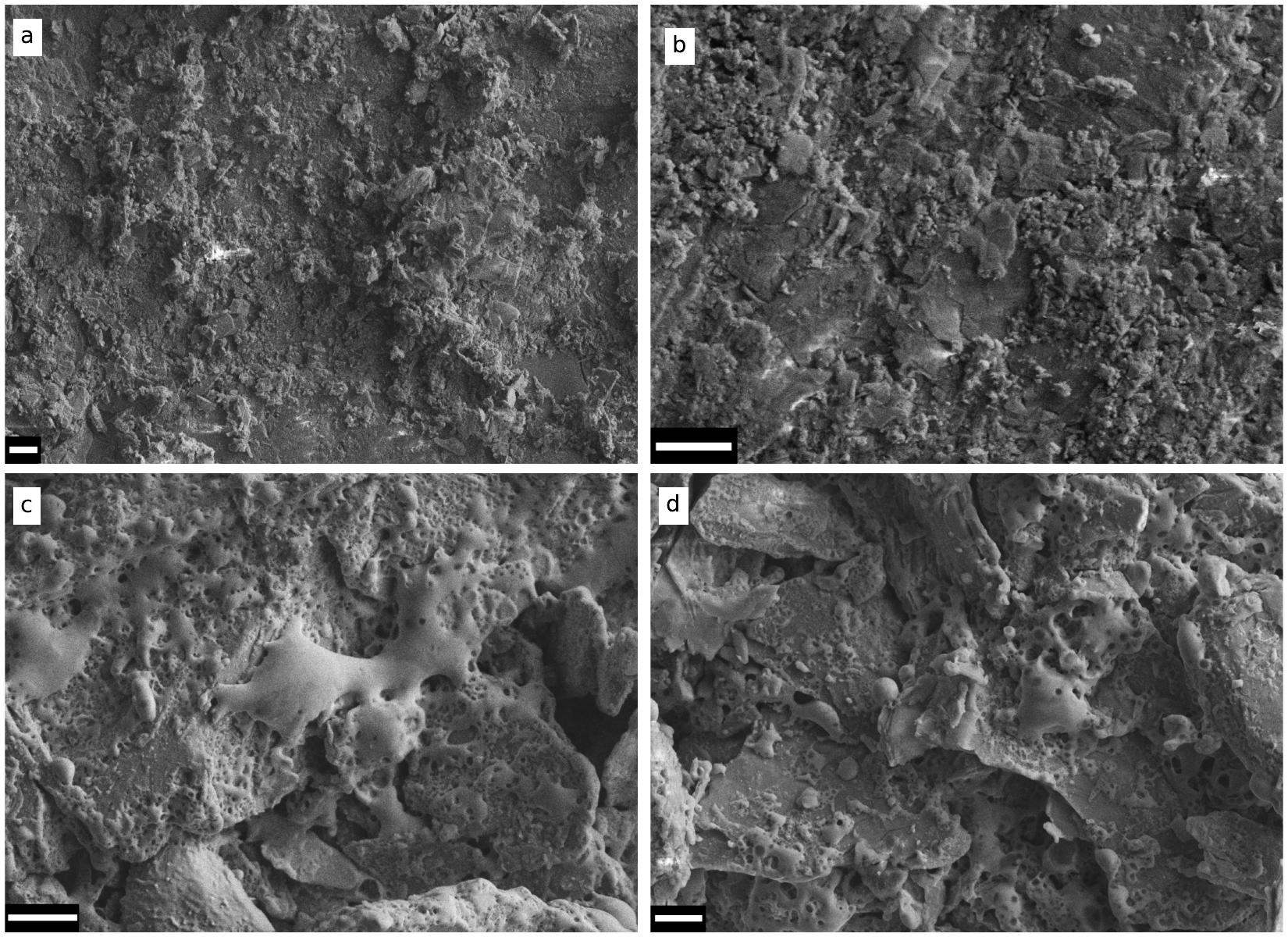}
    \caption{Secondary electron (SE) SEM images of a CI (a and c) and CM (b and d) simulant pellet before and after irradiation. Panels a and b show images of the fresh material, while panels c and d are of representative laser-irradiated regions.  The white scale bar at the lower left side of each panel corresponds to 2~${\mu}m$ in length.}
    \label{fig:sem}
\end{figure*}

\subsection{Microscopic Surface Morphology Analysis}

After laser irradiation and spectral analysis, we explored changes in the structural characteristics induced by laser irradiation for selected samples. In Figure 4, we compare a fresh surface of a selected CI and CM pellet to one laser irradiated with five pulses. Over the entire ablated surface, we find a ubiquitously distributed layer resembling melted material due to the fast heating induced after laser irradiation. The melted coating forms a web of features that are not identified in the unweathered surface. As seen in Figure \ref{fig:sem} (c-d), the altered surface exhibits a variety of irregular shapes and structures ranging from a few hundred nanometers in size (bubbles on the melts) up to ${\mu}m$ in length. Specifically, we find quasi-spherical vesiculated features ranging from 1 to 2-${\mu}m$ in diameter (see Figure \ref{fig:sem}d). Embedded to these melted splash web of features, voids and bubbled spaces are widespread on their surface, typically around a few hundred nanometers in diameter.

\subsection{XPS Analysis}

\begin{figure*}[ht!]
    \centering
    \includegraphics[width=0.9\textwidth]{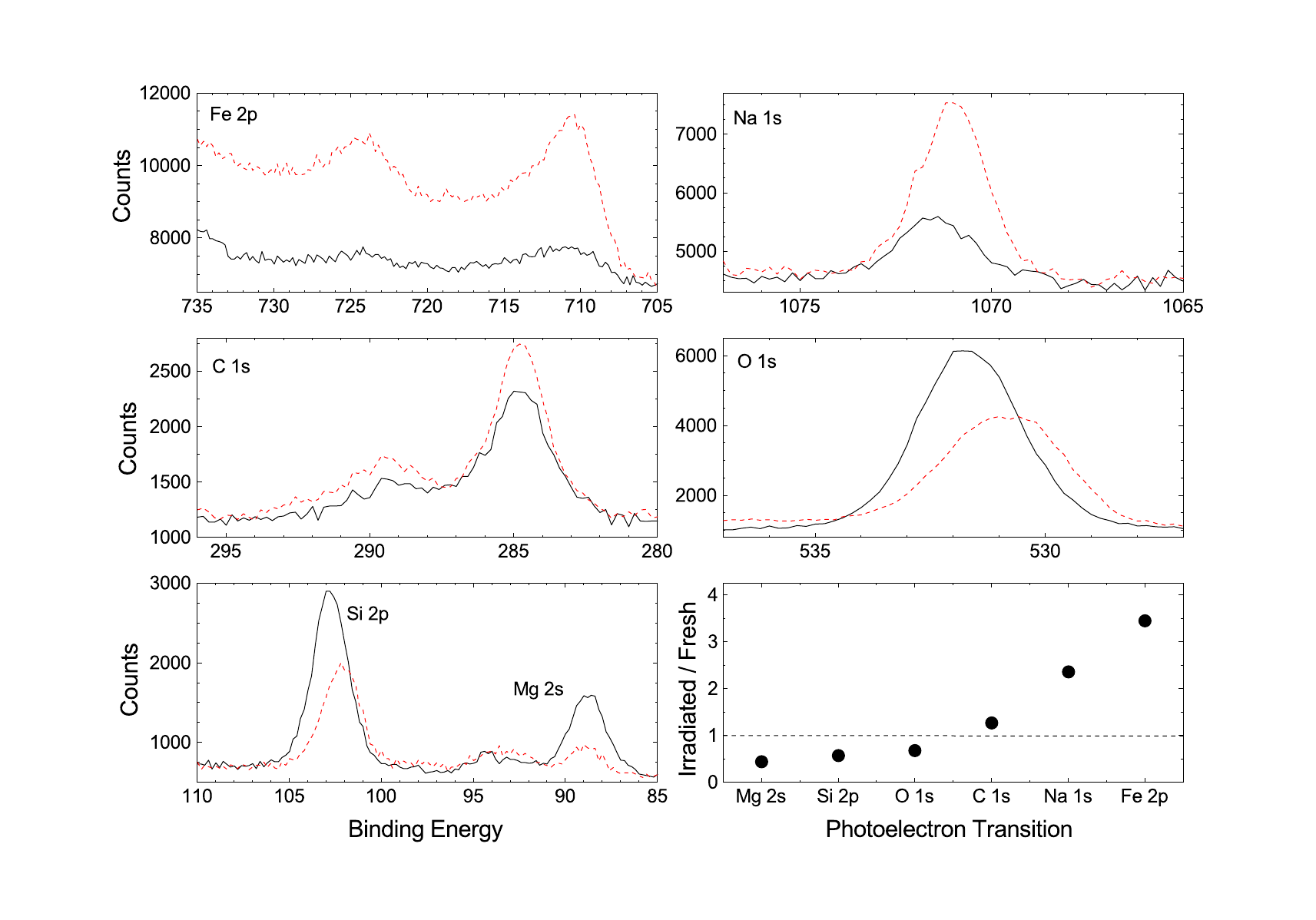}
    \caption{High-resolution XPS spectra of the fresh (black solid line) and weathered (red dashed line) CM sample. Bottom right: the ratio between the integrated area of the irradiated and the unirradiated surface for a given photoelectron transition.}
    \label{fig:xps}
\end{figure*}

We also performed ex-situ XPS analysis on select CM samples.  From a survey analysis of the unweathered surface, we find relative abundance variations of the studied elements to be less than 15\%. This heterogeneity is expected, given the simulant is composed of a number of different materials. Interestingly, we find that laser irradiation causes significant changes in the relative elemental abundances present on the surface of our sample. Figure \ref{fig:xps} shows the high-resolution spectra of weathered and fresh material taken from the same sample, along with the ratio of the band area of the main components after laser irradiation to the fresh surface. We find that the irradiated surface is enriched in C, Na, and Fe by 1.3$\times$, 2.4$\times$, and 3.5$\times$, respectively. On the other hand, the abundances of Mg, Si, and O are reduced by 0.4$\times$, 0.6$\times$, and 0.7$\times$, respectively. Even with ex-situ analysis, the compositional modification observed by XPS suggests that the altered layer observed in Figure \ref{fig:sem} is significantly different from the underlying grains.

\section{Discussion \label{sec:discussion}}

\subsection{Chemical and Spectral Alterations by Laser Irradiation}

\subsubsection{Compositional Changes due to Laser Irradiation}
SEM imaging of our irradiated sample shows the formation of melted material, splash particles ranging from $>$50~nm to ${\mu}m$ sized, and vesiculated-globular structures (see Figure \ref{fig:sem}). The features are consistent with melted structures found in previous studies \citep{Matsuoka_2020, Thompson2019, THOMPSON2020113775} and are ubiquitously distributed over the laser-irradiated surface. Based on previous TEM studies \citep{Thompson2019, CHAVES2023115634}, we assume the modification extends to at least 100~nm below the surface. Thus, as XPS only probes the top few nm of a sample surface \citep{Seah1979}, we conclude that the coating is preferentially enriched in Fe, C, and Na compared to the fresh underlying material. Based on the abundances of our simulant, the most likely minerals are magnetite and olivine (Fe) and coal (C). The Na likely originates from contaminants in the terrestrial minerals and, in the case of the CM simulant, from sodium metasilicate, which was used as a binder. However, we suspect the binder played a limited role in any of the observed changes, as both the CI and CM results (Figure \ref{fig:spectra} - \ref{fig:sem}) are essentially the same, and the preparation of the CI simulant did not require a binder \citep{Britt2019}. Below, we discuss the observed changes in spectral reflectance in the context of our SEM and XPS results. 

\subsubsection{Changes in Spectral Slope, Albedo and Absorption Features Caused by Laser Irradiation \label{sec:spectral_changes}}

We find that our initial reflectance spectrum for both samples has a blue slope in both the near-infrared and ultraviolet-visible regions and that laser irradiation causes the spectral slope to increase (become redder). After five laser pulses, the slope in the near-infrared is still slightly blue or nearly flat, while in the ultraviolet-visible is red, showing a change that is about 4x higher than was observed in the near-infrared. The increase in slope with laser irradiation is consistent with our previous mid-infrared studies of these simulants \citep{Prince2022} and is similar to what has been seen in other laboratory experiments on carbon-bearing materials but not others (see section \ref{sec:intro}). Interestingly, our observed trends appear to be opposite to what is predicted by \citet{LANTZ201810} in that our CI/CM samples do not brighten/turn blue but darken/turn red with laser irradiation. However, we point out that the initial albedo of our CI/CM simulant sample pellets is about 3-5 times higher than their CI/CM samples and is similar to their other meteorite samples (CO/CV) that reddened and darkened in response to space weathering \citep{Lantz2017}.  Thus, although composition is important for determining any sample’s albedo, it may be that the initial albedo of the sample is most important in determining the weathering trends in carbonaceous meteorite samples.

We speculate that the spectral reddening observed in our experiments is caused by the observed coating on our sample surface. Recent TEM analysis of Ryugu samples \citep{Noguchi2023, Matsumoto2024}, as well as of laser irradiated carbonaceous chondrite samples \citep{Matsuoka2015, Matsuoka_2020, Thompson2019, THOMPSON2020113775}, show an abundance of nanophase products containing sulfur. However, if those products are formed in our experiments, we suspect that they do not play an important role in altering the reflectance spectra of our samples, as the CI simulant contains 2.5 – 3.3 times more sulfur than the CM simulant (as estimated from wt\% analysis of initial samples and follow-up with XRF \citep{Britt2019}), yet the reflectance changes caused by laser irradiation in both CI and CM simulants are essentially the same (Figure \ref{fig:spectra} - \ref{fig:brightness}). Thus, given that iron shows the largest enhancement in XPS analysis, it seems likely it plays an important role. For instance, we have seen that laser-produced deposits of olivine contain chemically reduced iron, as well as npFe \citep{Loeffler2008}, and so those phases are anticipated in our samples as well. Chemically reduced iron has been shown to both redden and darken reflectance spectra for a number of minerals \citep{Hapke2001, Sasaki2001, BRUNETTO2006546, Loeffler2008}. We did not observe this chemical reduction with XPS, because we had to remove the sample from the vacuum chamber for analysis, which will cause the chemically reduced iron within the depth probed by XPS to reoxidize rapidly \citep{Loeffler2009, Loeffler2016}. Future Transmission Electron Microscopy analysis may help to identify the presence of reduced iron phases, including npFe, as these reduced phases are often found to be preserved below depths than we are probing with XPS \citep{Loeffler2016, Thompson2019}.  

In addition to the slope changes, we also observed that both samples darkened after a single laser pulse. Darkening is consistent with what was seen for Murchison powders \citep{Matsuoka2015, Matsuoka_2020}, as well as chipped samples and laser deposits \citep{Thompson2019, THOMPSON2020113775}, and is commonly observed in this spectral region for laser irradiated silicate minerals \citep{Yamada1999, Sasaki2001, BRUNETTO2006546, Loeffler2016}. While chemically reduced iron could also be driving the darkening observed here, other factors could contribute. For instance, the SEM images (Figure \ref{fig:sem} (c-d)) show voids and melted structures (i.e., bubbles, splash particles, and vesiculated structures in Figure \ref{fig:sem}) produced by the laser are also on the order of the size of the UV-NIR wavelengths, and these types of structural changes may also act to lower the albedo of the sample \citep{BRUNETTO2006546, Matsuoka_2020}. We note that \citet{Hiroi1999} and \citet{Matsuoka_2020} suggested that the amorphization and destruction of phyllosilicates could also cause the darkening of carbonaceous chondrites materials. While destruction of the phyllosilicates in the 3~${\mu}m$ region has been observed in a previous study \citep{Noguchi2023}, we did not observe any decrease in the 3~${\mu}m$ absorption band during laser irradiation of these samples \citep{Prince2022}. Thus, although it seems unlikely that dehydration of our sample occurred, the altered region of the sample is likely amorphous \citep{Loeffler2008, Noguchi2023}. 

Besides the initial darkening, we observed that subsequent pulsing caused additional, albeit weaker, changes to the albedo: the shorter wavelengths continued to darken and the longer wavelengths brightened. Previously, the brightening in the visible region in response to ion irradiation has been attributed to compositional changes in organics present in asphaltite \citep{Moroz2004b}. This pathway was also given as a possibility for spectral changes in response to laser irradiation of Murchison \citep{THOMPSON2020113775}, which saw an increase in reflectance (after a subsequent drop) after multiple laser pulses; this increase was nearly equal in both the visible and near-infrared regions. Thus, in our case, it may be that both the chemically reduced iron, which will act more strongly at shorter wavelengths, and the processed carbon, which may act relatively equally at all wavelengths in our spectral range, are both contributing to the albedo changes after multiple laser pulses.   

Besides the changes in spectral slope and albedo in our samples, we observed the altered layer did not significantly change the features near 0.27~${\mu}m$ or 2.3~${\mu}m$, suggesting that space weathering would not be able to remove these features \citep{Hendrix2019}. Unlike these absorptions, the laser essentially removed the $\sim$0.6~${\mu}m$ feature after a single laser pulse. Although this feature is unlikely the absorption typically associated with hydration on phyllosilicates, as our serpentine is Mg-rich \citep{Britt2019} and the feature observed on airless bodies and in meteorite samples is believed to originate in Fe-bearing phyllosilicates (Fe$^{+2}$ to Fe$^{+3}$ charge transfer) \citep{Gaffey1976, CLOUTIS2011309}, its location is fortuitous and suggests that space weathering may have a strong effect on the detectability of any feature in this region, including the 0.7~${\mu}m$ feature. In the future, we hope to perform direct tests on samples containing Fe-bearing phyllosilicates to test this hypothesis.

\begin{figure}[ht!]
    \centering
    \includegraphics[width=0.47\textwidth]{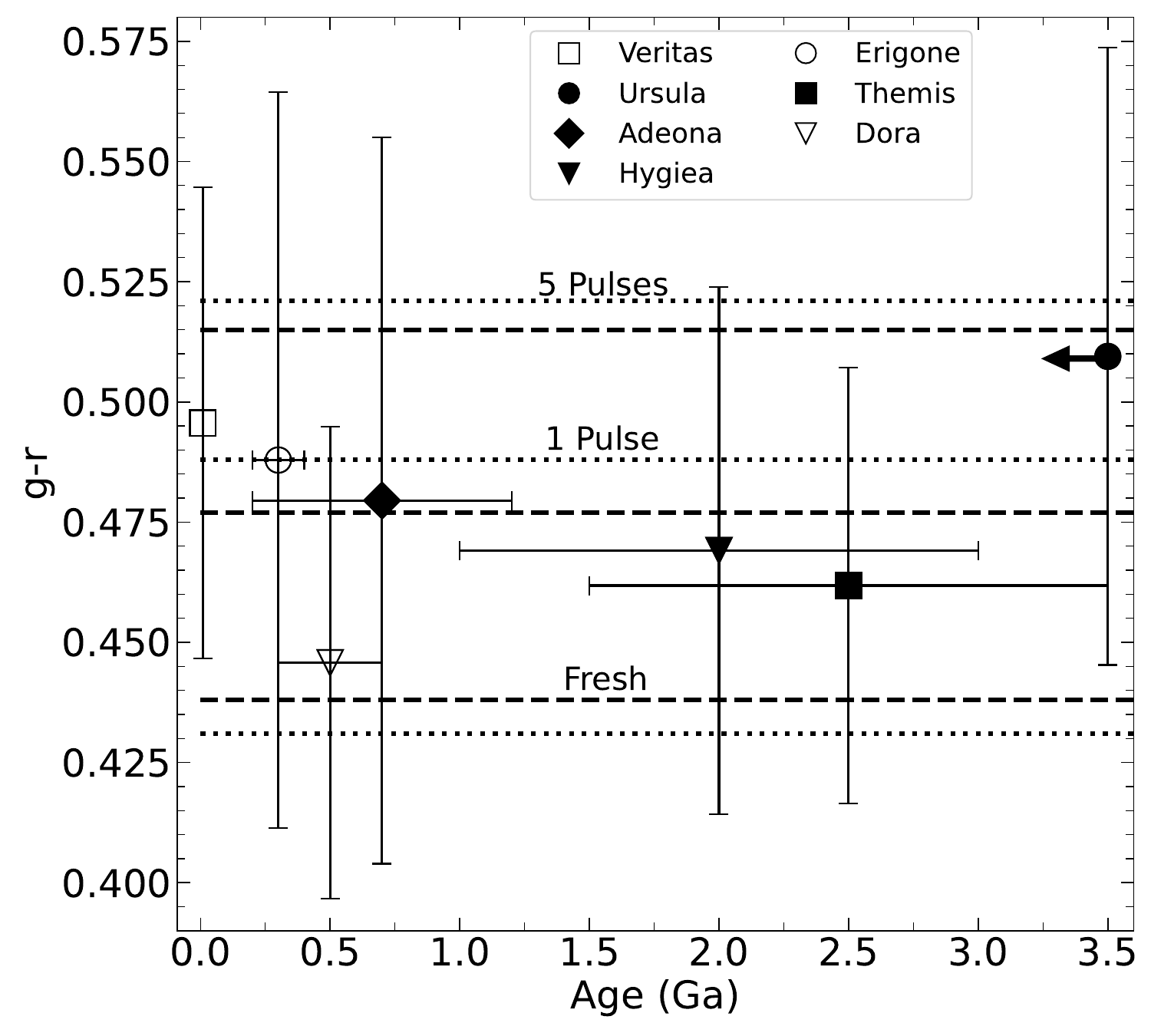}
    \caption{SDSS $g-r$ color of C-complex asteroid families as a function of their dynamical ages. The error bars of the family colors correspond to the standard deviation among the whole members. The horizontal dotted and dashed lines indicate the $g-r$ colors of the CI and CM laboratory spectra, respectively. Each set of lines corresponds to laboratory spectra after 0, 1 and 5 laser pulses.}
    \label{fig:timescale}
\end{figure}

\subsection{Astrophysical implications \label{sec:astrophysical}}

Analysis of our laboratory data (spectral slope, albedo, and absorption features) suggests that the ultraviolet-visible region appears to be more sensitive to space weathering than the near-infrared region, as has been suggested for S-class asteroids previously \citep{Hendrix2006TheEO, Brunetto2015}. Thus, in the following, we focus on comparing our results to remote sensing data taken at near-ultraviolet to visible wavelengths. 

In Figure \ref{fig:timescale}, we compare the $g-r$ color for each of the C-complex families considered here with the derived color for our fresh and laser-weathered CI and CM pellets. The color of the family corresponds to the average of the color distribution for every family member, while the error bars represent the square root of the average of the squared deviations from the mean. We do not observe an explicit trend between the color of families and their dynamic age, as younger families (e.g. Veritas and Erigone) share similar colors with older ones (Hygiea and Themis). However, very few of these families do include members that have colors bluer (i.e., $g-r$ $<$ 0.43 mags) than the color of our unweathered material, but instead, most of them have optical colors similar to our laser-weathered sample.

\begin{figure}[ht!]
    \centering
    \includegraphics[width=0.47\textwidth]{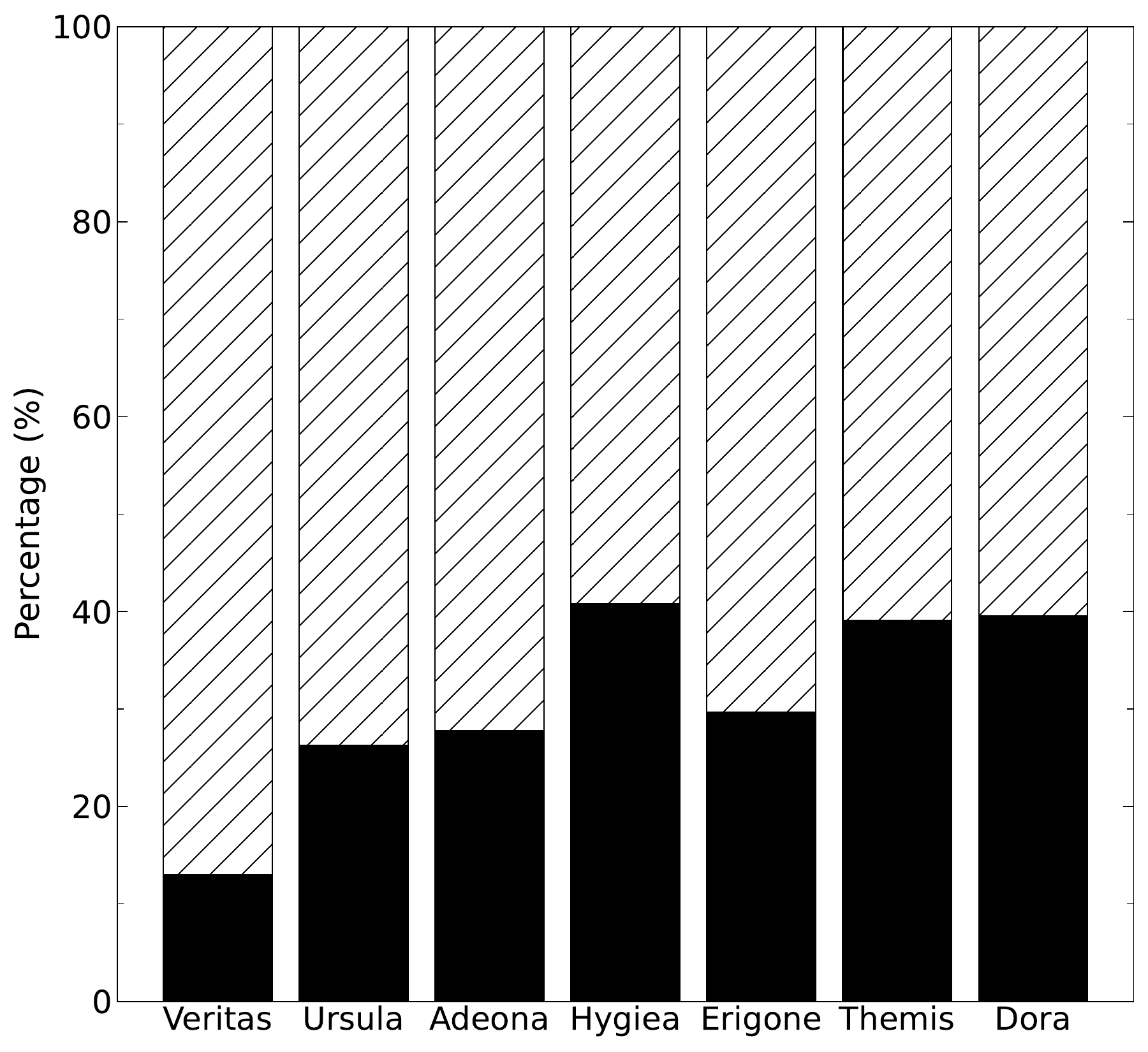}
    \caption{Distribution of the SDSS \textit{griz} visible spectral slope of C-complex asteroid families. The filled bars correspond to the percentage of blue (negatively sloped) asteroids in a given family, while the dashed bars correspond to the percentage of red (positively sloped) asteroids.}
    \label{fig:slope_percent}
\end{figure}

In addition to examining the average photometric color over an entire family, we also measured the spectral slope distribution in the visible region (i.e., converting the SDSS-g, -r, -i, and -z magnitudes into reflectance) within each family (see Figure \ref{fig:slope_percent}), defining any object with a negative spectral slope as blue or a positive spectral slope as red. Interestingly, nearly 70\% of the total C-complex population considered in this analysis are spectrally red, suggesting that most carbonaceous surfaces populating these primitive families possess spectrally red (positively sloped) surfaces in the visible region between 0.37 and 0.72~${\mu}m$.

The observation that the spectral properties of the majority of C-complex families are most similar to our space-weathered samples suggests that the weathering timescale is near-equal or faster than the time it takes to refresh the surface regolith. Previous estimates for resurfacing on a 5~km object initiated by impacts, which create global seismic waves that act to destabilize the near-surface of the asteroid regolith (i.e., ``vertical launching''; \cite{RICHARDSON2005325}) is $\sim$10$^{6}$~yr \citep{RIVKIN20111294}. Assuming a similar time estimate for C-complex families, this resurfacing time is slightly longer than the time estimated for micrometeorite impacts when considering a large mass range of hypervelocity impactors \citep{Loeffler2009}, although, admittedly, both estimates are fairly uncertain. We note that the weathering timescale may be shorter if one also considers the solar wind, which has been estimated to act on a much shorter timescale than micrometeorite impacts \citep{Loeffler2009, Chrbolkova2021} and should cause spectral alteration like what we have observed if the initial albedo of the samples is similar (see section \ref{sec:spectral_changes}). While determining the relative importance of these two alteration mechanisms is difficult, the similarities seen in electron microscopy images of laser-irradiated Murchison samples \citep{Matsuoka_2020, Thompson2019} and in Ryugu samples \citep{Noguchi2023} suggests that at the very least, micrometeorite impacts on carbonaceous asteroids are important. Nonetheless, a weathering timescale shorter than 10$^{6}$~yr would also be consistent with observations of the Veritas family located at the outer regions of the main belt (3.17~AU). Veritas has been estimated to only be 8.3~My old \citep{Nesvorny2015}, yet 87\% of the objects have a red slope.

In addition, the 0.7~${\mu}m$ spectral feature (typically referring to an absorption band between 0.5 to 0.8~${\mu}m$) was detected in asteroids by \citet{Villas1989} and associated with aqueous alteration in phyllosilicates \citep{Lipschutz1999, Nakamura2006}. Interestingly, many observational surveys have identified the 0.7~${\mu}m$ feature on carbonaceous asteroids families \citep{VILAS1994456, hiroi1996, CARVANO2003356, Mothe-diniz2005, CLOUTIS2011309, Howell2011, Fornasier2014, Morate2016, Depra2020} although others have not \citep{Deleon2016,PINILLAALONSO2016}. Judging from the rapid disappearance of the 0.6~${\mu}m$ broad feature in our experiments, we suspect that space weathering may inhibit the ability to detect any feature in this spectral region, including the 0.7~${\mu}m$. As an example, Erigone, a $\sim$300~My \citep{Nesvorny2015} old family located at 2.3~AU, was found to be populated with around 58\% hydrated primitive asteroids (i.e., B-, X-, T-, and C-types; \citet{Morate2016}). Thus, it seems possible that most objects in this family experience resurfacing on a time scale shorter than weathering, thus yielding the detection of the 0.7~${\mu}m$ feature.

Finally, the Fe$^{2+}$ charge transfer UV feature at 0.27~${\mu}m$ has been identified in some dark asteroids, such as 702 Alauda, 704 Interamnia, and 41 Daphne \citep{Hendrix2019}. Our experiments support these detections, as we find that this feature is not strongly affected by space weathering. Future laboratory studies expanding the low wavelength extreme of this absorption feature may be useful to more quantitatively assess whether this absorption feature shows minor variations with space weathering.

\section{Conclusion}

We performed pulsed-laser irradiation of CI and CM simulants to simulate progressive space weathering induced by micrometeorite bombardment. We find that laser irradiation causes an increase in spectral slope (reddening) across the UV-VIS-NIR range wavelengths and a decrease in the albedo (darkening), and these changes are significantly stronger in the ultraviolet-visible region. We speculate that these changes are driven by the excess iron that is produced in the altered surface region on our sample surface, but other factors, such as structural changes in the sample surface, may also contribute. In addition, although the 0.27~${\mu}m$ band appears relatively stable under laser irradiation, a broad feature at 0.6~${\mu}m$ rapidly disappears with laser irradiation, suggesting that space weathering may inhibit the detection of any feature in this spectral region, including the 0.7~${\mu}m$ band, which has typically been used an indicator of hydration. Furthermore, the observed spectral changes appear to be consistent with our analysis of a large dataset of asteroids from the SDSS MOC, which shows that the majority of the objects appear to be spectrally red in the visible, while also possessing colors that are more similar to our irradiated material than our fresh samples. Finally, we also find that ``younger'' and ``older'' C-complex families have similar colors, suggesting that the space weathering process is near-equal or faster than the time it takes to refresh the surfaces of these airless bodies. 

\acknowledgements 

\ This material is in part supported by the National
Science Foundation Graduate Research Fellowship Program under grant No. 2021318193 to ALO. Any opinions, findings, conclusions, or recommendations expressed in this material are those of the author(s) and do not necessarily reflect the views of the National Science Foundation.
\ MJL acknowledges funding from NASA's Solar Systems Workings program. 
\ This research has made use of NASA's Astrophysics Data System.
\ We thank Ronald S. Allen for his helpful contribution to the SEM analysis at the Northern Arizona University Imaging and Histology Core Facility.
\ We thank the students Beau Prince and Emily Clark for their support with the laser irradiation system.

Funding for the Sloan Digital Sky Survey V has been provided by the Alfred P. Sloan Foundation, the Heising-Simons Foundation, the National Science Foundation, and the Participating Institutions. SDSS acknowledges support and resources from the Center for High-Performance Computing at the University of Utah. The SDSS web site is \url{www.sdss.org}.

SDSS is managed by the Astrophysical Research Consortium for the Participating Institutions of the SDSS Collaboration, including the Carnegie Institution for Science, Chilean National Time Allocation Committee (CNTAC) ratified researchers, the Gotham Participation Group, Harvard University, Heidelberg University, The Johns Hopkins University, L’Ecole polytechnique f\'{e}d\'{e}rale de Lausanne (EPFL), Leibniz-Institut f\"{u}r Astrophysik Potsdam (AIP), Max-Planck-Institut f\"{u}r Astronomie (MPIA Heidelberg), Max-Planck-Institut f\"{u}r Extraterrestrische Physik (MPE), Nanjing University, National Astronomical Observatories of China (NAOC), New Mexico State University, The Ohio State University, Pennsylvania State University, Smithsonian Astrophysical Observatory, Space Telescope Science Institute (STScI), the Stellar Astrophysics Participation Group, Universidad Nacional Aut\'{o}noma de M\'{e}xico, University of Arizona, University of Colorado Boulder, University of Illinois at Urbana-Champaign, University of Toronto, University of Utah, University of Virginia, Yale University, and Yunnan University. 

\software{Astropy \citep{AstroPy2013},
          Scipy \citep{scipy2020},
          Numpy \citep{numpy2020},
          Matplotlib \citep{matplotlib2007},
          Pathfinder X-ray Microanalysis Software\footnote{\url{https://www.thermofisher.com/order/catalog/product/IQLAADGABKFAQOMBJE}}, CASA XPS \citep{FAIRLEY2021100112}}

\bibliography{sample63}{}
\bibliographystyle{aasjournal}

\end{document}